\documentclass[12pt]{article}

\begin{document}

\title{New Equations for Gravitation with the Riemann Tensor\\ and 4-Index
Energy-Momentum Tensors for Gravitation and Matter\thanks{%
Published in: ''Zakir Z. (2003) Structure of Space-Time and Matter. CTPA,
Tashkent''.}}
\author{Zahid Zakir\thanks{E-mail: zahid@in.edu.uz}\\Centre for Theoretical Physics and Astrophysics,\\ P.O.Box 4412, Tashkent 700000 Uzbekistan}
\date{May 3, 1999;\\
Revised October 17, 2003}
\maketitle

\begin{abstract}
A generalized version of the Einstein equations in the 4-index form,
containing the Riemann curvature tensor linearly, is derived. It is shown,
that the gravitational energy-momentum density outside a source is
represented across the Weyl tensor vanishing at the 2-index contraction. The
4-index energy-momentum density tensor for matter also is constructed.
\end{abstract}

\section{Introduction}

The definition of the energy-momentum for the gravitational field is a more
complicated procedure then for matter \cite{Ei}. Due to the vanishing of the
Einstein tensor $G_{ik}$ outside a source:

\begin{equation}
G_{ik}=\frac{1}{\kappa }(R_{ik}-\frac{1}{2}g_{ik}R)=0,
\end{equation}
one can not treat it directly as the energy-momentum density for gravitation.

The standard methods of asymptotically flat space-time become meaningful
only at large distances from the source. Particularly, the introduction of
pseudotensors leads to difficulties which have not been overcame. Even the
such rigorous treatment of the gravitational energy as the Hamiltonian
formulation leads to the conclusion about its non-localizability due to
non-tensor character of their observables (see \cite{Fa}).

At the same time, the curvature tensor $R_{iklm}$ more than all other
observables can be considered as a true characteristics of gravity. For this
reason, the explanation of the statement about non-localizability of the
gravitational energy by the principle of equivalence is incorrect because of
the local and covariant character of the curvature tensor, appeared in the
theory also due to that principle.

In the present paper a generalized version of the Einstein equations
directly containing the Riemann tensor will be formulated. Really, the
tensor $G_{ik}$, vanishing in empty space, in fact is the contraction of the
nonvanishing 4-index tensor: 
\begin{equation}
g^{il}G_{iklm}=G_{km}=0.
\end{equation}
In the paper we shall derive $G_{iklm}$ from the Einstein-Hilbert action
function as:

\begin{equation}
G_{iklm}=\frac{1}{\kappa }\left[ R_{iklm}-\frac{1}{2(d-1)}%
(g_{il}g_{km}-g_{im}g_{kl})R\right] ,  \label{G}
\end{equation}
where $d$ is the spacetime dimensionality, and then obtain 4-index
gravitational equations:

\begin{equation}
G_{iklm}=T_{iklm}.
\end{equation}
Here the 4-index energy-momentum density tensor of the source $T_{iklm}$ is
defined as: 
\begin{equation}
T_{iklm}=V_{iklm}+T_{iklm}^{(m)},
\end{equation}
where $T_{iklm}^{(m)}$ is appropriately symmetrized a combination of the
standard energy-momentum density tensor of matter $T_{ik}$ and its scalar $T$%
. Here $V_{iklm}$ is the new truly 4-index energy-momentum density tensor
with the property $g^{il}V_{iklm}=0$, which does not vanish outside the
source (in the vacuum) and, therefore, can be interpreted as {\it the
energy-momentum density of the gravitational field}.

In the vacuum $G_{iklm}$ contains only the Weyl tensor $C_{iklm}$ and we
have the equations for the gravitational field:

\begin{equation}
\frac{1}{\kappa }C_{iklm}=V_{iklm}.
\end{equation}

\section{The action function and field equations with the Riemann tensor}

The gravitational equations we obtain from the Einstein-Hilbert action
function:

\begin{equation}
S=\int d\Omega \sqrt{-g}\left( -\frac{1}{2\kappa }R+L\right) ,
\end{equation}
where $L$ is the matter Lagrangian. The variation of the geometric term can
be represented as:

\begin{equation}
\delta S_{g}=-\frac{1}{4\kappa }\delta _{g}\int d\Omega \sqrt{-g}%
(g^{il}g^{km}-g^{im}g^{kl})R_{iklm}.
\end{equation}
The relationships:

\begin{equation}
g^{il}\delta R_{il}=g^{il}\delta (g^{km}R_{iklm})=g^{il}\delta
g^{km}R_{iklm}+g^{il}g^{km}\delta R_{iklm},
\end{equation}
allows one rewrite the variations of $\delta R_{iklm}$ in the form: 
\begin{equation}
g^{il}g^{km}\delta R_{iklm}=-g^{il}\delta g^{km}R_{iklm}+g^{il}\delta R_{il}.
\end{equation}

Then, by taking into account the formulae: 
\begin{eqnarray}
\delta (\sqrt{-g})R &=&-\frac{1}{2}\sqrt{-g}\delta g^{km}g_{km}R= \\
&=&\sqrt{-g}\delta g^{km}g^{il}\left[ -\frac{1}{2(d-1)}%
(g_{km}g_{il}-g_{kl}g_{im})R\right] ,
\end{eqnarray}
we obtain:

\begin{equation}
\delta S_{g}=-\frac{1}{2\kappa }\int d\Omega \sqrt{-g}(G_{iklm}g^{il})\delta
g^{km},
\end{equation}
where $G_{iklm}$ is presented in Eq.(\ref{G}).

The Riemann tensor can be represented as: 
\begin{eqnarray}
R_{iklm} &=&C_{iklm}+\frac{1}{(d-2)}%
(g_{km}R_{il}-g_{kl}R_{im}+g_{il}R_{km}-g_{im}R_{kl})+  \nonumber \\
&&-\frac{1}{(d-1)(d-2)}(g_{il}g_{km}-g_{im}g_{kl})R,
\end{eqnarray}
where $d$ is the spacetime dimensionality, and $C_{iklm}$ has the property $%
g^{il}C_{iklm}=0$. If $C_{iklm}=0$, the such manifold is conformally flat.

We introduce a corresponding 4-index energy-momentum density tensor for the
source with the symmetry properties the same as for $R_{iklm}$ as: 
\begin{equation}
T_{iklm}=V_{iklm}+T_{iklm}^{(m)}.
\end{equation}
Here $V_{iklm}$ is the truly 4-index part of the energy-momentum density
tensor of the source with the vanishing contraction, and $T_{iklm}^{(m)}$ is
constructed from the 2-index energy-momentum density tensor of matter $%
T_{km} $ as:

\begin{eqnarray}
T_{iklm}^{(m)} &=&\frac{1}{(d-2)}%
(g_{km}T_{il}-g_{kl}T_{im}+g_{il}T_{km}-g_{im}T_{kl})-  \nonumber \\
&&-\frac{T}{(d-1)(d-2)}(g_{il}g_{km}-g_{im}g_{kl}), \\
T &=&g^{km}T_{km}=\frac{1}{2}(g^{il}g^{km}-g^{im}g^{kl})T_{iklm}.
\end{eqnarray}

Then, for the variation of source's action function we have:

\begin{equation}
\delta _{g}S_{m}=\frac{1}{2}\int d\Omega \sqrt{-g}\delta
g^{km}g^{il}T_{iklm}.
\end{equation}
The result of the variational procedure, therefore, is:

\begin{equation}
\delta S=-\frac{1}{2}\int d\Omega \sqrt{-g}\delta g^{km}g^{il}\left(
G_{iklm}-T_{iklm}\right) .
\end{equation}
which gives the field equations:

\begin{equation}
g^{il}(G_{iklm}-T_{iklm})=0.
\end{equation}

Therefore, we obtain 4-index equations for the gravitational field in the
form:

\begin{equation}
G_{iklm}=T_{iklm},
\end{equation}
or:

\begin{equation}
\frac{1}{\kappa }\left[ R_{iklm}-\frac{1}{2(d-1)}(g_{il}g_{km}-g_{im}g_{kl})R%
\right] =T_{iklm}.
\end{equation}

The covariant derivatives of these 4-index tensors in the case $d=4$ are: 
\begin{eqnarray}
G_{.klm;i}^{i} &=&\frac{1}{\kappa }\left[ R_{.klm;i}^{i}-\frac{1}{6}%
(g_{km}R_{,l}-g_{kl}R_{,m})\right] =  \nonumber \\
&=&T_{km;l}-T_{kl;m}-\frac{1}{3}(g_{km}T_{,l}-g_{kl}T_{,m}), \\
T_{klm;j}^{j(m)} &=&\frac{1}{2}\left[ T_{km;l}-T_{kl;m}-\frac{1}{3}%
(g_{km}T_{;l}-g_{kl}T_{;m})\right] =\frac{1}{2}G_{.klm;i}^{i}.
\end{eqnarray}
Then we obtain the relationship:

\begin{equation}
V_{klm;j}^{j}=G_{klm;j}^{j}-T_{klm;j}^{j(m)}=\frac{1}{2}G_{.klm;i}^{i}.
\end{equation}
and, therefore, 
\begin{equation}
V_{klm;j}^{j}=T_{klm;j}^{j(m)}
\end{equation}

In the vacuum, therefore, there are local conservation laws:

\begin{equation}
G_{\cdot klm;j}^{j}=V_{klm;j}^{j}=0.
\end{equation}

\section{Conclusions}

The very important part of the Riemann tensor is the Weyl tensor $C_{iklm}$
which, in fact, defines the gravitational field outside the source. This
part of the Riemann tensor ordinarily has been lost at the 2-index
contraction and this was a reason for the difficulties with the definition
of the gravitational energy.

It is shown that the 4-index energy-momentum tensor for matter must contain
the additional pure 4-index term $V_{iklm}$ which has all required
properties of the energy-momentum tensor for the gravitational field. The
discussion of some applications of this new treatment of the gravitational
energy and its connections with other definitions will be presented in \cite
{Za}.

\end{document}